\documentclass[12pt]{article}
\usepackage{graphicx}
\begin{document}
\begin{titlepage}

\title{Repulsive gravity near naked singularities and point massive particles}

\author{J. W. Maluf$\,^{(\ast)}$\\
Instituto de F\'{\i}sica, \\
Universidade de Bras\'{\i}lia\\
C. P. 04385 \\
70.919-970 Bras\'{\i}lia DF, Brazil\\}
\date{}
\maketitle
\begin{abstract}
We investigate the existence of repulsive gravitational acceleration near
naked singularities. The investigation is carried out by means of the
acceleration tensor, which is a coordinate invariant object. We find that the 
gravitational acceleration is repulsive in the vicinity of the origin in the 
Reissner-Nordstr\o m and in the Kerr space-times, and attractive at large
distances in the expected Newtonian way. We further address the space-time of
a point massive particle, which also exhibits repulsive effects near the origin.

\end{abstract}
\thispagestyle{empty}
\vfill
\par
\bigskip
\noindent $(\ast )$ wadih@unb.br, jwmaluf@gmail.com\par
\end{titlepage}
\newpage

\section{Introduction}
Under some general physical assumptions, the occurrence of singularities is an
inevitable feature of the exact solutions of Einstein's general relativity. The
existence of a singularity may be determined by the emergence of non-spacelike
geodesic incompleteness in the space-time as, for example, the sudden 
disappearance of a timelike geodesic (i.e., an observer) after a finite
amount of proper time (for a recent review, see Ref. 
\cite{SpaceSing}). In principle, singularities arise in the solutions of the
field equations in the form of black holes, or naked singularities. In the
vicinity of the singularities the energy densities, the space-time curvature and
torsion tensors, as well as any other geometrical field quantities, diverge in
powers of $1/r$. This is an indication that the usual description of the
space-time, as well as the ordinary laws of physics, break down 
\cite{SpaceSing}.

The existence of singularities in space-time is connected to one of the most
important problems in general relativity, which is the final fate
of the complete gravitational collapse of a massive body, such as a star.
Investigations on this issue have led to the conclusion that the gravitational 
collapse of several distinct matter field configurations may end up either as 
black holes or naked singularities (Ref. \cite{PRD104031} and references 
therein). Therefore, the latter are not ruled out as the final outcome of the 
physical process. If naked singularities exist in nature as real astrophysical
objects, it is important to establish conditions for distinguishing them from
black holes. Attempts in this direction have
been put forward in several investigations. The purpose is to find out 
observational differences in the space-time around black holes and naked
singularities by analysing: the circular motion of charged test
particles in the Kerr \cite{PRD044030} and Reissner-Nordstr\o m space-times
\cite{PRD104052}, the emergence of accretion disks \cite{PRD104031} and of 
gravitational lensing \cite{PRD063010,PRD103004,PRD124014}
in the strong field limit, around both types of 
singularities. The idea behind these analyses is to provide conditions for 
distinguishing naked singularities from black holes.

Repulsive gravity effects were conjectured to exist around naked singularities
in the Kerr \cite{PRD044030,Stuchlik,PRD104001} and Reissner-Nordstr\o m 
space-times \cite{PRD104052,PRD084023}. The repulsive effects
would explain discontinuities in the accretion disks around the singularities.
Although these repulsive effects are very speculative, they could 
exist in the vicinity of naked singularities if the latter are actual 
physical manifestations of general relativity. Repulsive gravitational effects 
have not been widely analysed in the literature. An investigation was carried 
out in Ref. \cite{Herrera} in the context of Weyl type metrics, by analysing 
the geodesic behaviour of test particles. A proposal for an invariant
definition of repulsive gravity, based on the properties of the Riemann tensor, 
was presented in Ref. \cite{Quevedo}.

In this article, we obtain the gravitational acceleration in the vicinity of
naked singularities and point massive particles
by means of the acceleration tensor. This tensor arises in the
analysis of reference frames in the space-time described by the metric tensor
$g_{\mu\nu}$ and by a set of tetrad fields $e^a\,_\mu$. The tetrad
fields yield the frame $e_a\,^\mu$ that is adapted to a field of observers in
space-time, defined by an arbitrary congruence of timelike world lines.
The acceleration tensor gives the values of the inertial (i.e., 
non-gravitational) accelerations that are necessary to maintain the frame in a
given inertial state  
(static, stationary or otherwise). If the frame is
maintained static in space-time, then the inertial acceleration
is exactly minus the gravitational acceleration imparted to the frame.

Although the procedure to be considered ahead is very simple, it has not been
presented in the literature so far. The procedure allows the determination of 
the emergence of repulsive gravitational acceleration on frames in arbitrary
space-times. Reference frames have been discussed in Refs. 
\cite{Maluf1,Maluf2}, in connection to the teleparallel equivalent of general
relativity \cite{Maluf3}. The acceleration tensor $\phi_{ab}=-\phi_{ba}$ is a
coordinate independent quantity, but depends on the frame ($a,b$ are $SO(3,1)$
indices). Linear accelerations and rotations are not absolute concepts. They
depend on a frame. The acceleration tensor may be used
to characterize the frame in space-time. The underlying geometrical structure
of this analysis is suitable to the teleparallel equivalent of general
relativity, where the gravitational field strength is the torsion tensor
$T_{a\mu\nu}$.

In Section 2 we recall the definition and properties of the acceleration tensor.
We will consider static or stationary observers in the point massive particle,
in the Reissner-Nordstr\o m and in the Kerr space-times in Section 3, where we
will describe the emergence of a repulsive acceleration near the origin in these
space-times. The Reissner-Nordstr\o m and Kerr space-times will display
naked singularities. Our notation is the following:
space-time indices $\mu, \nu, ...$ and SO(3,1) (Lorentz) indices
$a, b, ...$ run from 0 to 3. Time and space indices are indicated according to
$\mu=0,i,\;\;a=(0),(i)$. The tetrad fields are represented by $e^a\,_\mu$, and 
the torsion tensor by 
$T_{a\mu\nu}=\partial_\mu e_{a\nu}-\partial_\nu e_{a\mu}$.
The flat, tangent space Minkowski space-time metric tensor raises and lowers 
tetrad indices and is fixed by 
$\eta_{ab}= e_{a\mu} e_{b\nu}g^{\mu\nu}=(-1,+1,+1,+1)$.
The frame components are given by $ e_a\,^\mu $.

\section{The acceleration tensor}

The acceleration tensor is a generalization of the inertial acceleration 
4-vector of an observer defined by an arbitrary timelike world line $C$ in 
space-time. Let $C$ be denoted by $x^\mu(\tau)$, where the parameter $\tau$ is
the proper time of the observer. The velocity and acceleration of the observer 
along $C$ are denoted by $u^\mu=dx^\mu/d\tau$ and $a^\mu=Du^\mu/d\tau$, 
respectively. The absolute derivative $D/d\tau$ is constructed out of the
 Christoffel symbols $^0\Gamma^\mu_{\alpha\beta}$.

A frame adapted to the observer is constructed by identifying the timelike 
component of the frame $e_{(0)}\,^\mu$ with the velocity $u^\mu$ according to
$e_{(0)}\,^\mu=u^\mu(\tau)/c$ \cite{Hehl}. Throughout this article we will 
adopt $c=1$. Thus we have

\begin{equation}
a^\mu= {{Du^\mu}\over{d\tau}} ={{De_{(0)}\,^\mu}\over {d\tau}} =
u^\alpha \nabla_\alpha e_{(0)}\,^\mu\,.
\label{1}
\end{equation}
Out of the frame $e_a\,^\mu$ we may obtain the velocity and acceleration of
the observer on the timelike world line $C$. Therefore, a given set of tetrad 
fields, for which $e_{(0)}\,^\mu$ describes a congruence of timelike curves, is
adapted to observers characterized by the velocity field $u^\mu=e_{(0)}\,^\mu$
and acceleration $a^\mu$. The acceleration of the whole frame is 
determined by the absolute derivative of $e_a\,^\mu$ along $C$. Assuming that 
the observer carries an orthonormal tetrad frame $e_a\,^\mu$, the acceleration
of the frame along the path is given by \cite{Mashh2,Mashh3}

\begin{equation}
{{D e_a\,^\mu} \over {d\tau}}=\phi_a\,^b\,e_b\,^\mu\,,
\label{2}
\end{equation}
where $\phi_{ab}$ is the antisymmetric acceleration tensor. In analogy with the
Faraday tensor, we identify 
$\phi_{ab} \rightarrow ({\bf a}, {\bf \Omega})$, where ${\bf a}$ is the 
translational acceleration ($\phi_{(0)(i)}=a_{(i)}$) and ${\bf \Omega}$ is the 
frequency of rotation of the local spatial frame  with respect to a 
non-rotating, Fermi-Walker transported frame \cite{Mashh2,Mashh3}. 
It follows from Eq. (\ref{2}) that

\begin{equation}
\phi_a\,^b= e^b\,_\mu {{D e_a\,^\mu} \over {d\tau}}=
e^b\,_\mu \,u^\lambda\nabla_\lambda e_a\,^\mu\,.
\label{3}
\end{equation}

The acceleration vector $a^\mu$ defined by Eq. (\ref{1}) may be projected
on a frame in order to yield $a^b= e^b\,_\mu a^\mu=
\,e^b\,_\mu u^\alpha \nabla_\alpha e_{(0)}\,^\mu=\,\phi_{(0)}\,^b$.
Moreover, we find that

\begin{equation}
a^\mu= u^\alpha \nabla_\alpha e_{(0)}\,^\mu 
= {{d^2 x^\mu}\over {ds^2}}+\,\,^0\Gamma^\mu_{\alpha\beta}
{{dx^\alpha}\over{ds}} {{dx^\beta}\over{ds}}\,.
\label{4}
\end{equation}
We see that if $u^\mu=e_{(0)}\,^\mu$ represents a geodesic
trajectory, then the frame is in free fall and 
$a^\mu=0=\phi_{(0)(i)}$. Therefore we conclude that non-vanishing
values of the latter quantities represent inertial accelerations
of the frame.

After a number of simple manipulations, it is possible to rewrite
the acceleration tensor in the form \cite{Maluf1,Maluf2,Maluf3}

\begin{equation}
\phi_{ab}={1\over 2} \lbrack T_{(0)ab}+T_{a(0)b}-T_{b(0)a}
\rbrack\,,
\label{5}
\end{equation}
where $T_{abc}=e_b\,^\mu e_c\,^\nu T_{a\mu\nu}$, and 
$T_{a\mu\nu}=\partial_\mu e_{a\nu}-\partial_\nu e_{a\mu}$ is the 
torsion tensor of the Weitzenb\"ock space-time. The expression of $\phi_{ab}$
is invariant under coordinate transformations. 

The values of the 6 components of the acceleration tensor may
be used to characterize the frame, since $\phi_{ab}$ is not invariant under 
local SO(3,1) (Lorentz) transformations \cite{Maluf2}. Alternatively, the
frame may be characterized (i) by the identification $u^\mu=e_{(0)}\,^\mu$ (this 
equation fixes 3 components, because $e_{(0)}\,^0$ is
fixed by normalization), and (ii) by the 3 orientations in the 
three-dimensional space of the components 
$e_{(1)}\,^\mu,e_{(2)}\,^\mu,e_{(3)}\,^\mu$ \cite{Maluf1,Maluf3}.

For a frame that undergoes the usual translational and/or rotational 
accelerations in flat space-time, $\phi_{ab}$ yields the expected, ordinary
values \cite{Maluf1}. An interesting application
of the acceleration tensor is the following. Let us consider a static observer
in the Schwarzschild space-time, described by the standard Schwarzschild 
coordinates $(t,r,\theta,\phi)$.
The frame of the observer must satisfy $e_{(0)}\,^i=0=u^i$. The
non-vanishing components of the acceleration tensor form a vector 
${\bf a}= \phi_{(0)(1)} \hat{{\bf x}}+\phi_{(0)(2)} \hat{{\bf y}}
+\phi_{(0)(3)} \hat{{\bf z}}$, which eventually reads \cite{Maluf4}

\begin{equation}
{\bf a}= a(r) \hat{\bf r} ={m\over r^2}\biggl( 1-{{2m}\over r}\biggr)^{-1/2}\,
\hat{\bf r}\,,
\label{7}
\end{equation}
where 
$$\hat {\bf r}=\sin\theta \cos \phi \,\hat{\bf x} +
\sin \theta \sin \phi \,\hat{\bf y} + \cos\theta\, \hat{\bf z}\,,$$
and $\hat{{\bf x}}$, $\hat{{\bf y}}$, $\hat{{\bf z}}$ are the usual unit 
vectors in the asymptotically flat space-time.
The expression above represents the inertial acceleration necessary to maintain
the frame static in space-time. Therefore it exactly cancels the  
gravitational acceleration exerted on the frame. The inertial acceleration is
oriented along the positive direction of the radial vector $\hat{\bf r}$. 
It is well known that $-a(r)$ is the geodesic acceleration of a free 
body towards the black hole.

\section{Naked singularities and repulsive acceleration}
In this section we will consider two types of singularities that arise in the
exact solutions of Einstein's equations. The Reissner-Nordstr\o m and Kerr
space-times exhibit naked singularities when the physical parameters
of the solutions acquire certain values. We will also consider the 
space-time of the point massive particle recently investigated by Katanaev
\cite{Katanaev}. It was shown in the latter reference that the Schwarzschild
metric in isotropic coordinates is a solution of Einstein's equations with a 
$\delta$-type source at the origin. The metric is defined everywhere in $R^4$, 
except on the world line at $r=0$. The latter is not a naked singularity because
the space-time is geodesically complete at the origin
\cite{Katanaev}. The space-time of the point massive particle displays a very 
interesting repulsive gravitational effect, already noted by
Katanaev.

\subsection{The Reissner-Nordstr\o m naked singularity}

The Reissner-Nordstr\o m space-time, parametrized by the total mass $m$ and by
the charge $Q$, is described by the line element

\begin{equation}
ds^2=-\alpha^2 dt^2+{1\over \alpha^2}dr^2+r^2d\theta^2+
r^2\sin^2\theta\, d\phi^2\,,
\label{8}
\end{equation}
where 
$$\alpha^2 = 1-{{2m}\over r}+ {Q^2 \over r^2}\,.$$
A naked singularity exists if $\alpha^2>0$ everywhere in space. This
condition takes place if $Q^2>m^2$, which is assumed in this analysis. In
order to obtain the non-vanishing components of the acceleration tensor, we
have first to define the frame. We will establish a frame adapted to static
observers in space-time, characterized by the velocity field 
$u^\mu=(u^0,0,0,0)$. This frame is constructed by requiring the condition
$e_{(0)}\,^i= u^i=0$, which implies $e^{(k)}\,_0=0$. In $(t,r,\theta,\phi)$ 
coordinates, the tetrad fields $e_{a\mu}$ that satisfy these properties is

\begin{equation}
e_{a\mu}=\pmatrix{-\alpha&0&0&0\cr
0&\alpha^{-1}\sin\theta\,\cos\phi&r\cos\theta\,\cos\phi
&-r\sin\theta\,\sin\phi\cr
0&\alpha^{-1}\sin\theta\,\sin\phi&r\cos\theta\,\sin\phi
&r\sin\theta\,\cos\phi\cr
0&\alpha^{-1}\cos\theta&-r\sin\theta&0\cr}\,,
\label{10}
\end{equation}

The non-vanishing components of the acceleration tensor are
$\phi_{(0)(1)}, \phi_{(0)(2)}$ and $\phi_{(0)(3)}$. Definition (5) yields

$$\phi_{(0)(i)}=T_{(0)(0)(i)}=e_{(0)}\,^\mu e_{(i)}\,^\nu \,T_{(0)\mu\nu}\,,$$
where $i=1,2,3$.
Since the space-time is spherically symmetric, the orientation of the spatial
axes of the frame is arbitrary. After simple calculations we obtain
$\,\phi_{(0)(1)}= (\partial_1 \alpha) \sin\theta \cos \phi$, 
$\,\phi_{(0)(2)}= (\partial_1 \alpha) \sin\theta \sin \phi$, and
$\,\phi_{(0)(3)}= (\partial_1 \alpha) \cos \theta$.
In terms of the vector $\hat {\bf r}=\sin\theta \cos \phi \,\hat{\bf x} +
\sin \theta \sin \phi \,\hat{\bf y} + \cos\theta\, \hat{\bf z}$, we find

\begin{equation}
{\bf a}= \phi_{(0)(1)} \hat{{\bf x}}+\phi_{(0)(2)} \hat{{\bf y}}
+\phi_{(0)(3)} \hat{{\bf z}} 
= {1\over {\alpha r^2}} \biggl( m-{Q^2 \over r} \biggr) {\hat{\bf r}}\,.
\label{12}
\end{equation}

We see that for values of the radial coordinate $r$ such that $m-Q^2/r <0$,
or $r<Q^2/m$, the acceleration ${\bf a}$ is directed along the negative 
direction of the vector $\hat {\bf r}$. It means that for values of 
the radial coordinate $r$ such that $r<Q^2/m$, the inertial acceleration on
the frame has to be directed towards the origin of the coordinate system to
maintain the frame static. Therefore the gravitational 
acceleration is repulsive when 
$r<Q^2/m$, and diverges in the limit $r\rightarrow \infty$.
For very large values of $r$ such that $r>>m$ and $r>> Q$, we find
${\bf a}\cong m/r^2\,{\hat{\bf r}} $, as expected, since the geodesic 
acceleration is $- m/r^2\,{\hat{\bf r}} $ in this limit.

The right hand side of Eq. (\ref{12}) indicates that $r<Q^2/m$ establishes
a region of repulsion. This feature shares similarities with the outcome of
Ref. \cite{Quevedo}. In the latter reference, the region of repulsion in the
same space-time geometry, obtained by a totally different procedure, is given
by $0<r<2Q^2/m$.

\subsection{The Kerr naked singularity}

The Kerr space-time is characterized by the total mass  $m$ and by angular
momentum per unit mass $a$. In terms of the Boyer-Lindquist coordinates, the 
Kerr space-time is described by the line element

\begin{equation}
ds^2=
-{{\psi^2}\over {\rho^2}}dt^2-{{2\chi\sin^2\theta}\over{\rho^2}}
\,d\phi\,dt
+{{\rho^2}\over {\Delta}}dr^2 
+\rho^2d\theta^2+ {{\Sigma^2\sin^2\theta}\over{\rho^2}}d\phi^2\,,
\label{13}
\end{equation}
with the following definitions:

\begin{eqnarray}
\Delta&=& r^2+a^2-2mr\,, \nonumber \\
\rho^2&=& r^2+a^2\cos^2\theta\,,   \nonumber \\
\psi^2&=&\Delta - a^2 \sin^2\theta \,, \nonumber \\
\Sigma^2&=&(r^2+a^2)^2-\Delta a^2\sin^2\theta\,,\nonumber \\
\chi&=&2amr\,.
\label{14}
\end{eqnarray}
We assume that $a>m$. In this case, Eq. (\ref{13}) represents a naked
singularity. The tetrad fields yields the metric tensor according
to $e^a\,_\mu e^b\,_\nu \eta_{ab}=g_{\mu\nu}$. 
Since $e^a\,_\mu$ is a kind of square root of $g_{\mu\nu}$, it may
not be defined in every region of the three-dimensional space. For instance, 
in the case of a black hole, if we choose 
the observer to be static in space-time, then the tetrad field is
defined only in the region $r> r_+^*$, where 
$r_+^*=m+\sqrt{m^2-a^2\cos^2\theta}$ represents the external boundary
of the ergosphere of the Kerr black hole, because 
inside the ergosphere it is not possible to  maintain any observer in 
static regime. Inside the ergosphere, all observers are necessarily dragged 
in circular motion by the gravitational field.

In the context of the naked singularity here considered, we find that the 
set of tetrad fields that are everywhere defined in the three-dimensional space,
except at the origin $r=0$, is the same one considered in Ref. \cite{Maluf5}. 
These tetrad fields satisfy Schwinger's time gauge condition $e_{(i)}\,^0=0$, 
and are adapted to a field of observers that rotate in space-time under the 
action of the gravitational field. The four-velocity of observers
that are dragged by the gravitational field in circular motion reads

\begin{equation}
u^\mu(t,r,\theta,\phi)
={{\rho \Sigma}\over{(\psi^2\Sigma^2+\chi^2\sin^2\theta)^{1/2}}}
(1,0,0,{\chi \over {\Sigma^2}})\,,
\label{15}
\end{equation}
where all functions are defined in Eq. (\ref{14}). The quantity
$\omega(r)= -g_{03}/g_{33}=\chi/ \Sigma^2$
is the dragging velocity of inertial frames.

The tetrad fields that are adapted to observers whose four-velocities are 
given by Eq. (\ref{15}), i.e., for which $e_{(0)}\,^\mu =u^\mu$, 
and whose $e_{(1)}\,^\mu$, $e_{(2)}\,^\mu $ and $e_{(3)}\,^\mu $ components in
cartesian coordinates are oriented along the unit vectors
$\hat{\bf x}$, $\hat{\bf y}$, $\hat{\bf z}$, respectively, are given by

\begin{equation}
e_{a\mu}=\pmatrix{-A&0&0&0\cr
B\sin\theta\sin\phi
&C\sin\theta\cos\phi& D\cos\theta\cos\phi&-E\sin\theta\sin\phi\cr
-B\sin\theta\cos\phi
&C\sin\theta\sin\phi& D\cos\theta\sin\phi& E\sin\theta\cos\phi\cr
0&C\cos\theta&-D\sin\theta&0}\,,
\label{16}
\end{equation}
where

\begin{eqnarray}
A&=& {{(g_{03}g_{03}-g_{00}g_{33})^{1/2}}\over{(g_{33})^{1/2}}}=
{1\over (-g^{00})^{1/2} }=
{{ (\psi^2\Sigma^2+\chi^2\sin^2\theta)^{1/2} }\over {\rho \Sigma}}\,,
\nonumber \\
B&=&-{{ g_{03}}\over {(g_{33})^{1/2} \sin\theta}}={\chi\over{\rho\Sigma}}\,,
\nonumber \\ 
C&=&(g_{11})^{1/2}={\rho\over \sqrt{\Delta}}\,, \nonumber \\
D&=&(g_{22})^{1/2}=\rho\,, \nonumber \\
E&=& {{(g_{33})^{1/2}}\over {\sin\theta}}={\Sigma\over \rho}\,.
\label{17}
\end{eqnarray}
The tetrad fields above are asympotically flat. In Cartesian coordinates
they may be written as $e_{a\mu}\cong \eta_{a\mu}+ (1/2) h_{a\mu}$ in the 
asymptotic limit $r\rightarrow \infty$, where $h_{a\mu}=h_{a\mu}(1/r)$. Equation
(\ref{16}) constitutes the unique configuration that satisfies the 6 
conditions previously discussed, namely, the fixation of $e_{(0)}\,^i=u^i$ and
of the asymptotic orientation of the spatial components $e_{(i)}\,^\mu$.

We are interested in calculating only the translational acceleration
${\bf a}=\phi_{(0)(1)}\hat{\bf x} + \phi_{(0)(2)} \hat{\bf y}+
\phi_{(0)(3)}\hat{\bf z}$. In this case we have
$\phi_{(0)(i)}= g^{00} g^{11} e_{(0)0}\, e_{(i)1} T_{(0)01} +
g^{00} g^{22} e_{(0)0}\, e_{(i)2} T_{(0)02}$,
where $T_{(0)01}=\partial_1 A$, $T_{(0)02}=\partial_2 A$, and
$\partial_1=\partial_r$, $\partial_2=\partial_\theta$.
The evaluation of ${\bf a}$ is long but straightforward. We find that
the translational inertial acceleration on the frame given by Eq. (\ref{15}) 
is given by

\begin{equation}
{\bf a}={\sqrt{\Delta} \over{2\rho A^2}}\,\partial_1(A^2)\, \hat{\bf r}+
{1\over {2\rho A^2}}\,\partial_2 (A^2)\,\hat{\theta}\,,
\label{19}
\end{equation}
where $\hat{\theta}=\cos\theta \cos\phi\,\hat{\bf x}+\cos\theta \sin\phi\, 
\hat{\bf y}- \sin\theta\,\hat{\bf z}$, and 

\begin{eqnarray}
\partial_1 (A^2)&=&{{2(r-m)}\over{\rho^2}}-{{2r}\over{\rho^2}}\biggl(
{{\psi^2\Sigma^2+\chi^2\sin^2\theta}\over{\rho^2\Sigma^2}}\biggr)
+{{4\chi am}\over{\rho^2\Sigma^2}}\sin^2\theta \nonumber \\
&& -{\chi^2\over{\rho^2\Sigma^4}}\sin^2\theta \lbrack 
4r(r^2+a^2)-2a^2(r-m) \sin^2\theta\rbrack\,,
\label{20}
\end{eqnarray}

\begin{eqnarray}
\partial_2(A^2)&=&{{2\sin\theta\cos\theta}\over \rho^2}\biggr[
{\chi^2\over \Sigma^2}(1+a^2\sin^2\theta)- \nonumber \\
&&-a^2\biggl( 1-{\psi^2\over \rho^2}\biggr)+
{\chi^2\over \Sigma^4} \Delta\, a^2\sin^2 \theta\biggr]\,.
\label{21}
\end{eqnarray}

Expression (\ref{19}) provides the translational inertial acceleration
necessary to maintain the frame stationary in
space-time. The frame is adapted to an observer in circular motion around the 
$z$ axis, at a fixed radial coordinate $r$, and rotates according to Eq. 
(\ref{15}). The dependence of ${\bf a}$ on the coordinates $(r,\theta)$ is 
intricate, but it certainly displays repulsive character near $r=0$. 
By inspecting Eq. (\ref{19}), it is not difficult to conclude that in the 
limit $r\rightarrow 0$ we have

\begin{equation}
{\bf a}\longrightarrow \ \ \ -{m\over{a^2(\cos^2\theta)^{3/2}}}\hat{\bf r}\,.
\label{22}
\end{equation}
The limit above comes from the first term on the right hand side of 
Eq. (\ref{20}).
Since ${\bf a}$ is directed along the negative direction of  
$\hat{\bf r}$, the gravitational acceleration on the frame in the vicinity of 
the origin is repulsive, and is more intense as the observer approaches the
equatorial plane determined by $\theta=\pi/2$. At large distances ($r>>m$, 
$r>> a$), the gravitational field is attractive in the expected Newtonian way,
since ${\bf a}\cong m/r^2\,{\hat{\bf r}} $ is oriented in the positive direction
of ${\hat{\bf r}}$, and cancels the geodesic  acceleration $-{\bf a}$.

\subsection{The point massive particle}

The space-time of a point massive particle has been addressed by Katanaev 
\cite{Katanaev} as an exact solution of Einstein's equations, by 
writing the energy-momentum tensor in terms of a $\delta$ function of a 
point particle with support at the origin of the coordinate system. Katanaev
concluded that the Schwarzschild solution in isotropic coordinates represents 
the space-time of the point massive particle, which is described by the line 
element

\begin{equation}
ds^2=-\alpha^2 dt^2+\beta^2\lbrack dr^2+r^2(d\theta^2+
\sin^2\theta\,d\phi^2)\rbrack\,,
\label{23}
\end{equation}
where 

\begin{equation}
\alpha^2=\biggl( { { 1-{m \over {2r}}\over {1+{m\over {2r}}}}}\biggr)^2\,,
\ \ \ \ \ \ 
\beta^2= \biggl( 1+ {m\over {2r}}\biggr)^4\,.
\label{24}
\end{equation}
The parameter $m$ represents the mass of the point particle and arises in
the energy-momentum tensor.
Katanaev observed that in the range $0<r<r_\ast\;$, where 
$r_\ast=m/2$ is the Schwarzschild radius in isotropic coordinates, the 
gravitational field is repulsive, in contrast to being
attractive for $r>r_\ast$ We will arrive at the same conclusion in the 
following. 

The tetrad fields adapted to static observers in the point massive space-time,
in spherical coordinates, are given by,

\begin{eqnarray}
e_{a\mu}=\pmatrix{-\alpha&0&0&0\cr
0&\beta\sin\theta\,\cos\phi&\beta r\cos\theta\,\cos\phi
&-\beta r\sin\theta\,\sin\phi\cr
0&\beta\sin\theta\,\sin\phi&\beta r\cos\theta\,\sin\phi
&\beta r\sin\theta\,\cos\phi\cr
0&\beta\cos\theta&-\beta r\sin\theta&0\cr}\,.
\label{25}
\end{eqnarray}
The components of the translational acceleration 
${\bf a}=\phi_{(0)(1)}\hat{\bf x} + \phi_{(0)(2)} \hat{\bf y}+
\phi_{(0)(3)}\hat{\bf z}$ are $\phi_{(0)(i)}=T_{(0)(0)(i)}$. After simple 
calculations we obtain

\begin{eqnarray}
T_{(0)(0)(1)}&=& {1\over {\alpha \beta}}\sin\theta \cos \phi 
(\partial_1 \alpha)\,, \nonumber \\
T_{(0)(0)(2)}&=& {1\over {\alpha \beta}}\sin\theta \sin \phi 
(\partial_1 \alpha)\,,\nonumber \\
T_{(0)(0)(3)}&=& {1\over {\alpha \beta}}\cos\theta 
(\partial_1 \alpha)\,.
\label{26}
\end{eqnarray}
Altogether, these terms yield

\begin{equation}
{\bf a}={
{m \over r^2 } 
\over 
{\biggl( 1-{m\over {2r}}\biggr)\biggl(1+ {m\over {2r}}\biggr)^3}}
\hat{\bf r} \ \ \ \equiv \ \ \ a(r)  \hat{\bf r}\,.
\label{27}
\end{equation}
It is easy to verify that

\begin{eqnarray}
r>r_\ast &:&\; \ \ \ \ a(r)> 0\;, \nonumber \\
r<r_\ast &:&\; \ \ \ \ a(r)< 0 \;,\nonumber \\
r\rightarrow r_\ast +\epsilon &:&\; \ \ \ \ a(r) \rightarrow \infty 
\ \ \ \ \texttt{if} \ \ \ \ \epsilon \rightarrow 0\;, \nonumber \\
r\rightarrow r_\ast -\epsilon &:&\; \ \ \ \ a(r) \rightarrow -\infty 
\ \ \ \ \texttt{if} \ \ \ \ \epsilon \rightarrow 0\;, \nonumber \\
r\rightarrow 0 &:& \ \ \ \ a(r) \rightarrow 0\,.
\label{28}
\end{eqnarray}
The frame must undergo an inertial acceleration $a(r)< 0$ for $r<r_\ast$, in
order to be static in space-time. Therefore in the range $0<r<r_\ast$ the
gravitational acceleration on the frame is repulsive, and it is possible
to show that $a(r)$ vanishes for $r=0$. These results are in agreement with 
those obtained by Katanaev \cite{Katanaev}, and indicate a modification of 
the Newtonian attraction at short distances.

\section{Concluding remarks}
In this article we have shown that in the vicinity of the Kerr and
Reissner-Nordstr\o m naked singularities, and of the massive point particle,
the gravitational field is repulsive, in contrast to the current understanding
that gravity is everywhere attractive. This feature was already noted in the 
study of accretion disks around naked singularities \cite{PRD104052}.  
The analysis was carried out by means of the acceleration tensor presented in 
Section 2, which plays an important role in the teleparallel equivalent of 
general relativity \cite{Maluf3}. The acceleration tensor is a 
frame dependent quantity, since acceleration depends on the frame. However, it 
is invariant under coordinate transformations. 
In each of the three situations addressed in Section 3, we have obtained the
exact expression of the translational inertial acceleration on the frames, in
either static or stationary regime. The acceleration tensor provides the 
values of the inertial accelerations that are necessary to impart to the frame
to maintain it in a certain inertial state in space-time. Therefore, for static 
frames in space-time, the inertial
acceleration is precisely minus the gravitational acceleration. 

The condition for the emergence of repulsive gravitational acceleration is very
simple. One establishes the frame for static or stationary observers,
constructs the translational inertial acceleration ${\bf a}$,

\begin{equation}
{\bf a}=T_{(0)(0)(1)}\hat{\bf x} + T_{(0)(0)(2)}\hat{\bf y}
+T_{(0)(0)(3)}\hat{\bf z}\,,
\label{29}
\end{equation}
and check the sign of each quantity $T_{(0)(0)(i)}$. The gravitational
acceleration points in the opposite direction of $T_{(0)(0)(1)}\hat{\bf x}$,
$T_{(0)(0)(2)}\hat{\bf y}$ and $T_{(0)(0)(3)}\hat{\bf z}$. The positive or
negative sign of $T_{(0)(0)(i)}$ means that the gravitational acceleration is 
attractive or repulsive, respectively.

It is interesting to investigate the consequences of the results obtained here
in the analysis of the formation of accretion disks. In particular, the 
emergence of $\cos^2\theta$ in the denominator of Eq. (\ref{22}) is a curious
feature. As the radius of the circular orbit of the frames approaches the 
Kerr naked singularity on the equatorial plane, the repulsive
gravitational acceleration increases, and diverges only in the limit 
$r\rightarrow 0$. If naked singularities are not just 
hypothetical astrophysical objects, but manifestations of the physical reality,
this feature may play a role in the formation of accretion disks around the 
singularities. It has been conjectured the existence of powerful 
repulsive forces on the equatorial plane of Kerr-like geometries that display
naked singularities \cite{SpaceSing}, in contrast to the geodesic behaviour 
around black holes. It is expected that astrophysical observations will
distinguish naked singularities from black holes.

\end{document}